\documentclass[twocolumn,showpacs]{revtex4}
\usepackage{epsfig}
\usepackage{graphicx}
\usepackage{amsmath}
\usepackage{amsfonts}
\usepackage{upgreek}
\begin{document}

\preprint{APS/123-QED}

\title{Elastocapillarity: \\ Surface Tension and the Mechanics of Soft Solids}

\author{Robert W. Style}
\affiliation{%
Mathematical Institute, Oxford University, Oxford, UK
}%

\author{Anand Jagota}
\affiliation{%
Department of Chemical and Biomolecular Engineering and Bioengineering Program, Lehigh University, Bethlehem, USA
}%

\author{C.-Y. Hui}%
\affiliation{%
Department of Mechanical and Aerospace Engineering, Cornell University, Ithaca, USA}

\author{Eric R. Dufresne}%
\email[]{eric.dufresne@mat.ethz.ch}
\affiliation{%
Department of Materials, ETH Zurich, CH-8093 Zurich, Switzerland
}

%

%

\date{\today}

\begin{abstract}
It is widely appreciated  that surface tension can dominate the behavior of liquids at small scales.
Solids also have surface stresses of a similar magnitude, but they are usually overlooked. 
However, recent work has shown that these can play an central role in the mechanics of soft solids such as gels. 
Here, we review this emerging field.
We outline the theory of surface stresses,  from both mechanical and thermodynamic perspectives, emphasizing the relationship between surface stress and surface energy.
We describe a wide range of phenomena at interfaces and contact lines where surface stresses play an important role.
We highlight how surface stresses causes dramatic departures from classic theories for wetting (Young-Dupr\'{e}), adhesion (Johnson-Kendall-Roberts), and composites (Eshelby).  
A common thread  is the importance of the ratio of surface stress to an elastic modulus, which defines a length scale below which surface stresses can dominate. 
\end{abstract}

\pacs{Valid PACS appear here}
\maketitle


\section{INTRODUCTION}
Soft solids are ubiquitous. 
They  include gels, creams, foams, 
rubbers, pressure sensitive adhesives and much of biological tissue.
Soft solids  have long been applied in cosmetics, adhesives, sealants, and padding. 
Furthermore, exciting new applications are developing in surgery, tissue engineering, flexible electronics and soft robotics 
(\emph{e.g.} \cite{shep11,mori12,wood13,suo99,ling11,kepl13,sun14,shin03}), often utilising the fact that soft solids can exhibit mechanical phenomena that differ qualitatively from hard engineering materials \cite{gord78}.
One key difference is that surface stresses, which play a minor role in the mechanics of stiff materials, can dominate the behavior of soft solids.

The importance of surface stresses on the mechanics of soft materials has only come to light in the last few years.
For example, surface stresses stabilize the surface of a soft solid slab \cite{chen12,jago12}, but break up soft solid filaments \cite{mora10}.
Surface stresses resist the deformation of fluid inclusions in a soft solid and stiffen fluid-solid composites \cite{styl15}.
Liquid droplets on soft substrates can violate the classic Young-Dupr\'{e} equilibrium \cite{styl12,marc12b,styl13}.
Stiff solid particles adhered to soft substrates do not obey the standard models for adhesive contacts  \cite{john71,carrillo2012contact,styl13c}.

The surface of a material has an energy penalty per unit area of surface, the \emph{surface energy}, $\gamma$ \cite{dege10}.
In liquids, the surface energy penalty gives rise to a tensile \emph{surface stress}, $\mathbf{\Upsilon}$, that opposes surface stretching. It
acts to minimise the surface to volume ratio of the liquid, causing small droplets to bead up, 
and larger volumes of liquid to have smooth, flat surfaces. 
Surface stress allows small dense objects to float at a liquid surface,  and is the source of the Laplace pressure difference across curved surfaces.
Generally, $\mathbf{\Upsilon}$ is a symmetric second order, two-dimensional tensor.
However, in simple liquids, $\mathbf{\Upsilon}$ is isotropic and thus can be represented by a scalar $\Upsilon$.
Another convenient property of simple liquids, as we shall see, is that $\gamma=\Upsilon$.
This has led to $\gamma$ and $\Upsilon$ being referred to interchangeably as the surface tension \cite{gibb06,dege10}.

Surface energies and stresses in solids are different in two key ways to their liquid counterparts.
First, the surface stress and surface energy are not generally equal,  
$\gamma\neq \Upsilon$, \cite{shut50}.
Therefore, one must use the term `surface tension' carefully.
Secondly,  solid surface stresses can be anisotropic and even compressive  \cite{nico55,gibb06,shut50,gurt75,gurt78,camm94}.

The relative importance of surface stress and bulk elasticity is a matter of length scale.
Just as a fluid's surface tension creates a jump in hydrostatic pressure across a curved interface (the Young-Laplace equation), 
surface stress causes a jump in the stresses across a solid interface.
For a solid surface with mean curvature $\cal K$ and isotropic, uniform surface stress $\Upsilon$, the jump in normal stress is simply $\Upsilon \cal K$.
This stress jump drives local deformation in the bulk of the solid.
For an elastic solid with Young's modulus $E$, stresses balance so that  $\Upsilon {\cal K}\sim \epsilon E$, where $\epsilon$ is the strain.
Thus, we expect significant deformations ($\epsilon\sim 1$) when ${1/\cal K} \sim  \Upsilon/E$.
This defines an \emph{elastocapillary length}, $\Upsilon/E$.
Generally, at scales much larger than $\Upsilon/E$, surface effects are negligible.
At  scales much smaller than $\Upsilon/E$, surface stresses dominate  and one observes dramatic departures from classic elastic behavior.

Elastocapillary phenomena at the continuum level are possible when $\Upsilon/E$ is much larger than  molecular scales.
The surface energies of soft materials are typically of order $10-100 ~\mathrm{mN}/\mathrm{m}$ \cite{dege10}.
Thus elastocapillarity can be very important in soft polymeric materials like gels  ($E\sim O( \mathrm{kPa}$))
and elastomers ($E\sim O(\mathrm{MPa})$).

We briefly note that there is a large body of prior work that has focussed on surface stresses in stiffer materials 
such as metals and other crystallographic solids (see previous reviews: \cite{camm94,spae00,hais01,sand03}).
In this review, we will focus on recent work that has evolved from new observations of surface stress effects in soft solids.

\section{ELASTOCAPILLARY PHENOMENA IN TWO-PHASE SYSTEMS}

We begin this section by deriving the fundamental boundary condition balancing bulk and surface stresses 
across an interface.  Then, we discuss the implications of this force balance in some examples involving two phases
separated by an interface.

\subsection{Surface stress as a  boundary condition at an interface}\label{ss:strbc}

\begin{figure}[htbp]
    \centering
    \includegraphics[width=6cm]{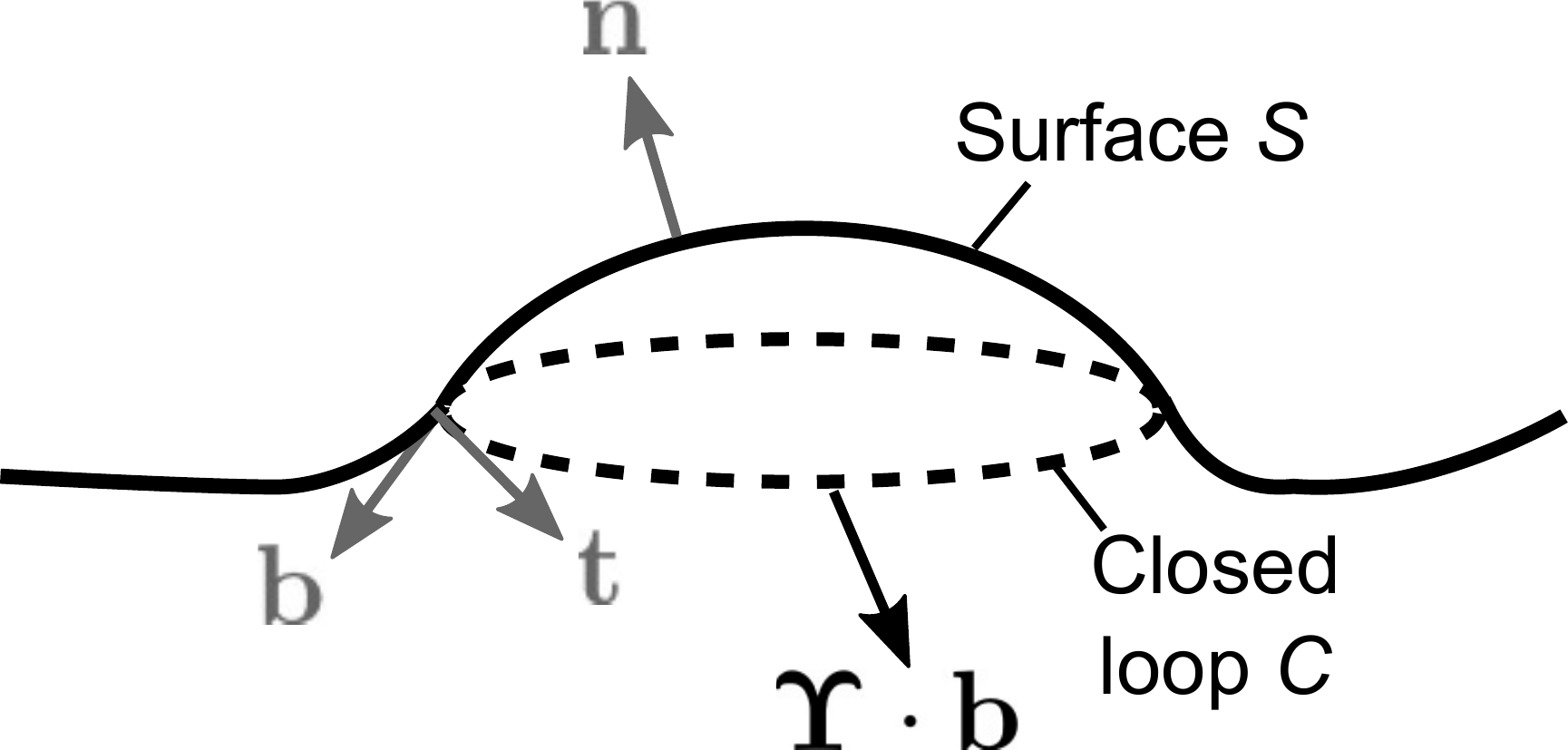} \\
    \caption{Schematic diagram for deriving the equation for force equilibrium at an interface.}
    \label{fig:forces}
\end{figure}

The surface stress tensor, $\mathbf{\Upsilon}$, is the 2D analogue of the more familiar 3D (Cauchy) stress tensor, $\boldsymbol{\sigma}$.
In 3D, the forces per unit area of a deformed surface (tractions) acting at a surface with normal $\mathbf{n}$ are given by $\boldsymbol{\sigma\cdot n}$.
The tractions can be resolved into a normal component, $\boldsymbol{n \cdot \sigma \cdot n}$, and a shear component, $\boldsymbol{t \cdot \sigma \cdot n}$, where 
$\mathbf{t}$ is a unit vector tangential to the surface.
Similarly, in 2D, the forces per unit length of a curve 
bounding a deformed surface with boundary normal $\mathbf{b}$ are given by $\mathbf{\Upsilon \cdot b}$ (see \textbf{Figure \ref{fig:forces}}).
Surface stresses always act in the plane of the surface and have a line-normal component, 
$\mathbf{b\cdot \Upsilon \cdot b}$, and tangential component, $\mathbf{t \cdot \Upsilon \cdot b}$, where $\mathbf{t}$ is the tangent vector to the curve.

The essential physics of elastocapillary coupling lies in a connection between the surface stress and the bulk stress at the interface.
The governing interfacial equation can be derived by considering the forces acting on the area of surface in \textbf{Figure \ref{fig:forces}}.
The condition for static equilibrium is,
\begin{equation}
\int_S\boldsymbol{\sigma_1}\cdot\mathbf{n}\,dS-\int_S\boldsymbol{\sigma_2}\cdot\mathbf{n}\,dS+\oint_C \mathbf{\Upsilon}\cdot \mathbf{b} \,dl=0.
\end{equation}
Here, the integrals in the deformed configuration and $\boldsymbol{\sigma_{1,2}}$ are the true (Cauchy) stresses in the deformed body. 
Note that we have ignored any forces acting on the surface due to external potentials (\emph{e.g.} surface charges in an electric field), and have assumed that the surface has no bending rigidity.
The first two terms represent forces due to bulk stresses acting on either side of the area of interface, $S$.
The last term is the contribution to total force due to surface stress. 
Using the surface divergence theorem, we can convert this term to a surface integral, and then remove the integrals (as $S$ is arbitrary) 
to obtain the generalisation of Laplace's law:
\begin{equation}
(\boldsymbol{\sigma_1}-\boldsymbol{\sigma_2}) \cdot \boldsymbol{n}=-\boldsymbol{\nabla^s}\cdot \mathbf{\Upsilon},
\label{eqn:YL1}
\end{equation}
where, $\mathbf{\nabla^s}$ is the surface gradient operator \cite{gurt75,gurt78,chen06}.
When the surface stress is isotropic, $\mathbf{\Upsilon}=\Upsilon\mathbf{I_2}$, where $\mathbf{I_2}$ is the 2D identity tensor, 
this boundary condition simplifies to
\begin{equation}
(\boldsymbol{\sigma_1}-\boldsymbol{\sigma_2}) \cdot \mathbf{n}=-\Upsilon {\cal K} \mathbf{n} + \mathbf{\nabla^s}\Upsilon,
\label{eqn:YL2}
\end{equation}
where curvature ${\cal K}=\mathbf{\nabla^s}\cdot\mathbf{n}$.
Note that the stresses due to the first term on the right-hand side are always normal to the interface.
The stresses due to the second are always in-plane, and equivalent to Marangoni stresses in a liquid \cite{javili2014unified}.
If the surface stress is both isotropic and uniform, $\mathbf{\nabla^s}\Upsilon=0$, and
\begin{equation}
\label{eq:gyl}
(\boldsymbol{\sigma_1}-\boldsymbol{\sigma_2}) \cdot \mathbf{n}=-\Upsilon {\cal K} \mathbf{n}.
\end{equation}
which reduces to the Young-Laplace law for fluid interfaces ($\Delta P = \gamma \cal K$).
The simplified boundary condition, Eq. (\ref{eq:gyl}) shows excellent agreement with a wide range of 
experimental observations in hydrogels and silicone gels (e.g. \cite{mora10,mora11,jeri11,mora13,styl13,styl13c,styl15,karp15}).
The implications for more complex surface stresses are just beginning to be explored \cite{andr15}.

\subsection{Surface Stresses Smooth Slabs}
\begin{figure}[htbp]
    \centering
    \includegraphics[width=8.cm]{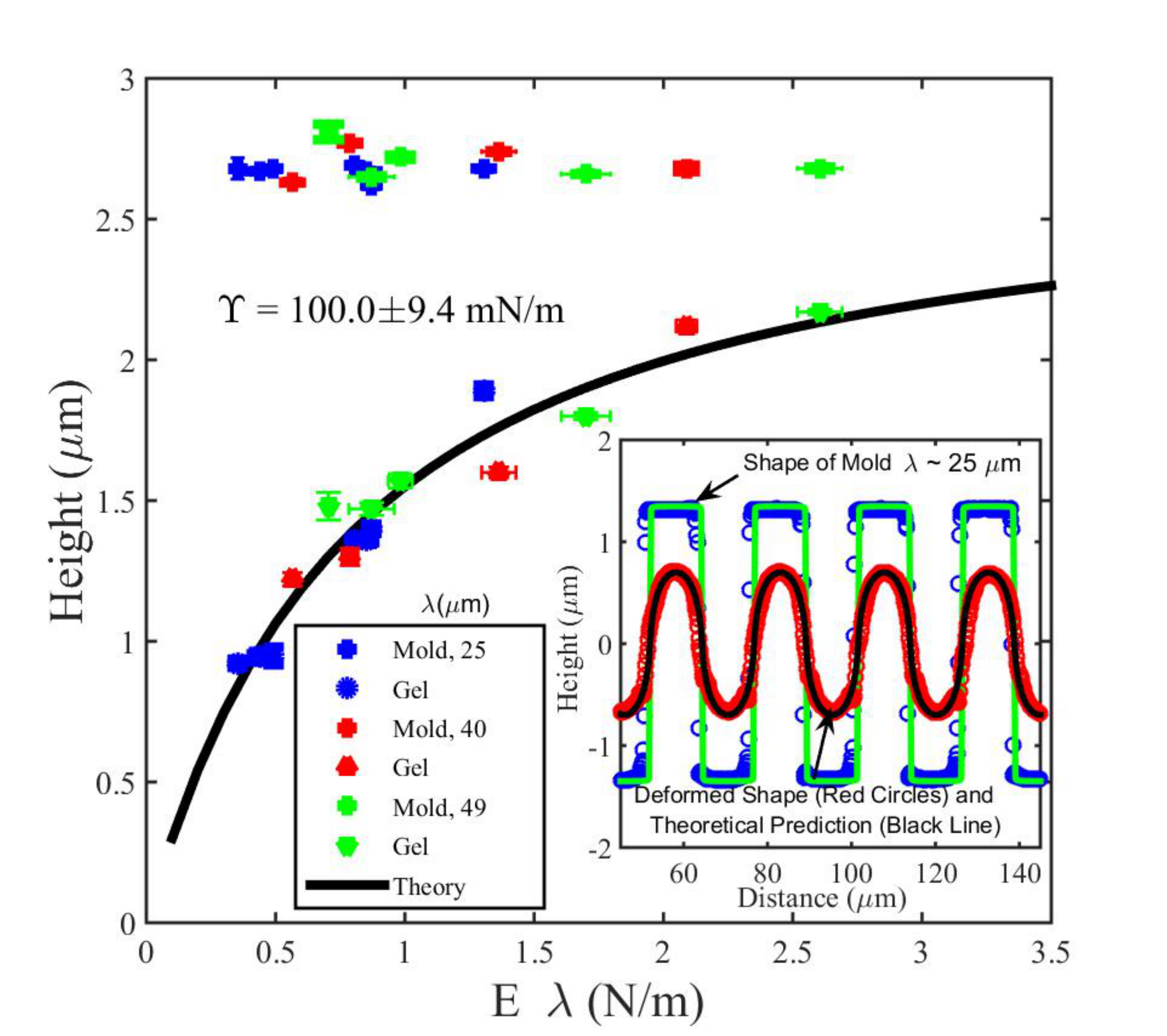} \\
    \caption{Flattening of a solid surface by surface stress. 
    A gelatin gel with Young's modulus $E=32.5$kPa is cured (stress-free) in a PDMS mould, then released and allowed to relax under the 
    influence of surface stress.Height of the surface profile before and after release 
    for different stiffnesses and wavelengths. The x-axis is Young's modulus times surface wavelength, 
    and the black curve shows the theoretical prediction \cite{pare14}. 
     Inset shows optical profilometry (theory) measurements of the surface profile of the mould: 
    blue (green) and after release: red (black).  
    }
    \label{fig:flattening}
\end{figure}

The simplest manifestation of the competition between surface stress and elasticity is the smoothing of features on solid slabs 
\cite{hui02,gord08,jago12,pare14,mora13}.
An example of this is the change in shape of soft gels that that are released from rigid, patterned molds.
\textbf{Figure \ref{fig:flattening}} (inset) shows the profile of a gel before and after release from a mold with a square wave pattern.
Upon release, sharp corners are rounded out and the overall amplitude of the pattern is reduced.
Flattening is strong for softer gels and shorter wavelengths (\textbf{Figure 
\ref{fig:flattening}}).

The flattening process is intimately linked to the elastocapillary length, $\Upsilon/E$ \cite{hui02,jago12}.
Consider a simplified experiment with a solid having an initially sinusoidal surface profile of 
wavelength $\lambda$ and amplitude $A$ \cite{jago12}.
According to Eq. \ref{eq:gyl}, the capillary stress normal to the surface is 
$\sigma_\Upsilon=\Upsilon {\cal K}$, where ${\cal K}\sim A/\lambda^2$ is the surface curvature. 
This stress drives flattening, and is opposed by the elastic stress.
Complete flattening requires the strain in the solid $\epsilon \sim A/\lambda$, 
so the elastic stress $\sigma_{el}\sim E \epsilon\sim EA/\lambda$.
Thus,  flattening is significant when $\sigma_\Upsilon\sim \sigma_{el}$, or when $\lambda \sim \Upsilon/E$. 
In short, elasticity will prevent deformation at longer length scales, and surface stress overwhelms elasticity and 
flattens the surface at smaller scales.
Similar arguments show that sharp corners will round out to leave a smooth surface with a 
radius of curvature $\sim \Upsilon/E$ \cite{hui02,gord08}.

\begin{figure*}[htbp]
    \centering
    \includegraphics[width=16cm]{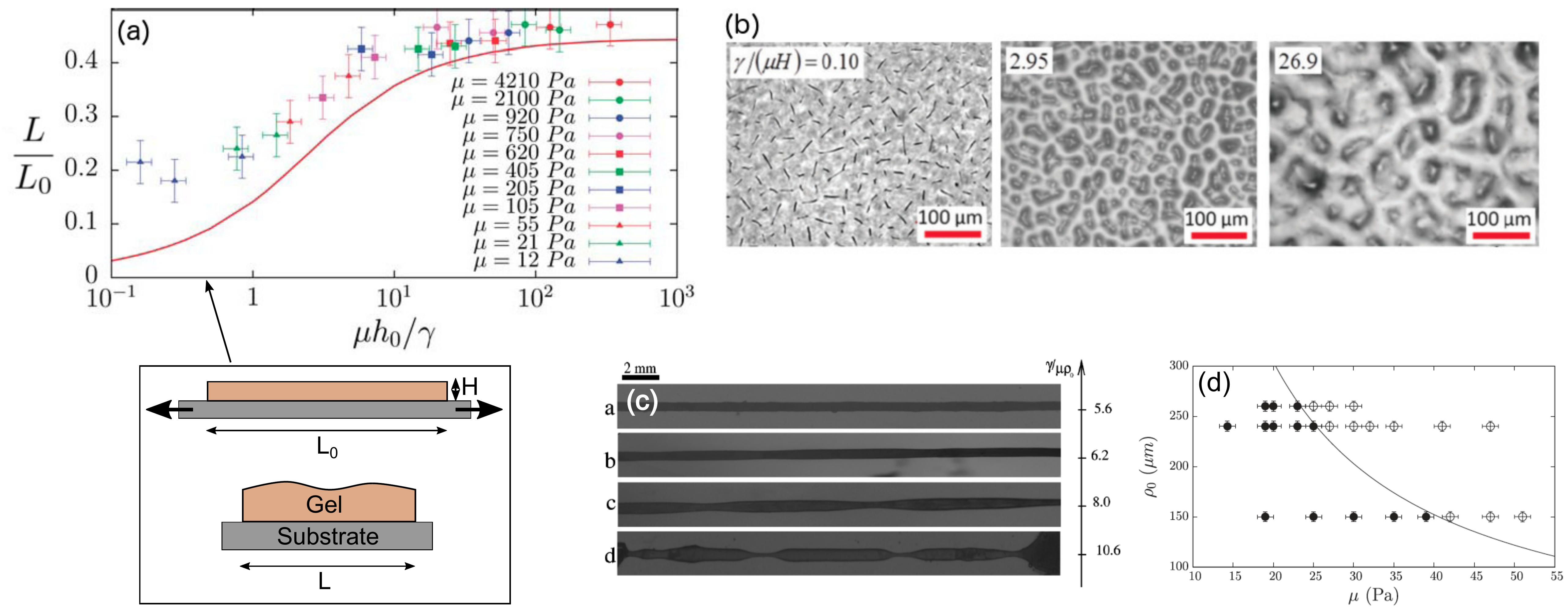} \\
    \caption{a) In uniaxial compression of a soft layer (see schematic), the critical compression for instability 
    depends only on the ratio $\Upsilon/H E$ \cite{mora11}. (b) Soft silicone layers of varying 
    thickness are swollen by applying an electric field across them. As the ratio $\Upsilon/H E$ varies, 
    there is a dramatic shift in the form of the surface instability \cite{wang13}. c) Cylinders of soft agar gels are released in toluene and allowed to relax. From top to bottom, the cylinders have shear moduli, $\mu$, reducing from 27 to 12Pa. d) The cylinders become unstable (black points) when their diameters ($2\rho_0$) are less than $\Upsilon/E$ --  this theoretical prediction is given by the curve. Unfilled points show stable cylinders \cite{mora10}. All these soft solids are approximately incompressible so $E\approx 3\mu$.}
    \label{fig:instab}
\end{figure*}

In a similar vein, surface stresses also suppress compressional instabilities of soft solids films.
It is well known that confined layers of elastic solids become unstable to buckling, wrinkling or creasing 
when compressed or swollen (e.g. \cite{wang14}).
However the morphology of the surface instability, the critical compression threshold, and the critical defect 
size for instability nucleation all depend sensitively on the relative size of the layer thickness, $H$, 
and elastocapillary length $\Upsilon/E$ (e.g. \textbf{Figure \ref{fig:instab}a,b}).
For example, when $\Upsilon/H E\ll 1$, the surface becomes unstable to creases with wavelength $\propto H$ \cite{mora11,chen12,wang13}.
However, when $\Upsilon/H E \gg 1$, the surface tends to wrinkle instead with wavelength $\gg H$, while the compression needed for the instability to occur increases significantly beyond classical predictions \cite{biot63,mora11,wang13,chak13,chau15}.

\subsection{Surface Stresses Deform Rods}

Some of the most dramatic effects of surfaces stress in solids are seen in objects with slender sections. 
Surface stresses deform long, thin cylindrical beams of soft hydrogels when they are released from rigid moulds \cite{mora10,mora13}.
Sufficiently thin cylinders develop undulations along their axis, as shown in \textbf{Figure \ref{fig:instab}c}  \cite{mora10}.
This instability is analogous to the famous Rayleigh-Plateau capillary instability \cite{plat73,rayl78}.
In that case, surface tension destablizes a liquid cylinder of radius $R$ to surface perturbations that have 
wavelength $\gtrsim O(R)$ .
In the case of a soft gel, however, this only occurs when the cylinder diameter is 
less than $\Upsilon/E$, as shown in \textbf{Figure \ref{fig:instab}c,d}.
Additionally, surface stresses will round out a slender object's corners, and when it has an asymmetric cross-section, the resulting asymmetric surface forces can cause significant bending \cite{mora13}.

\subsection{Surface Stresses Stiffen Inclusions and  Composites}

\begin{figure*}[htbp]
    \centering
    \includegraphics[width=16cm]{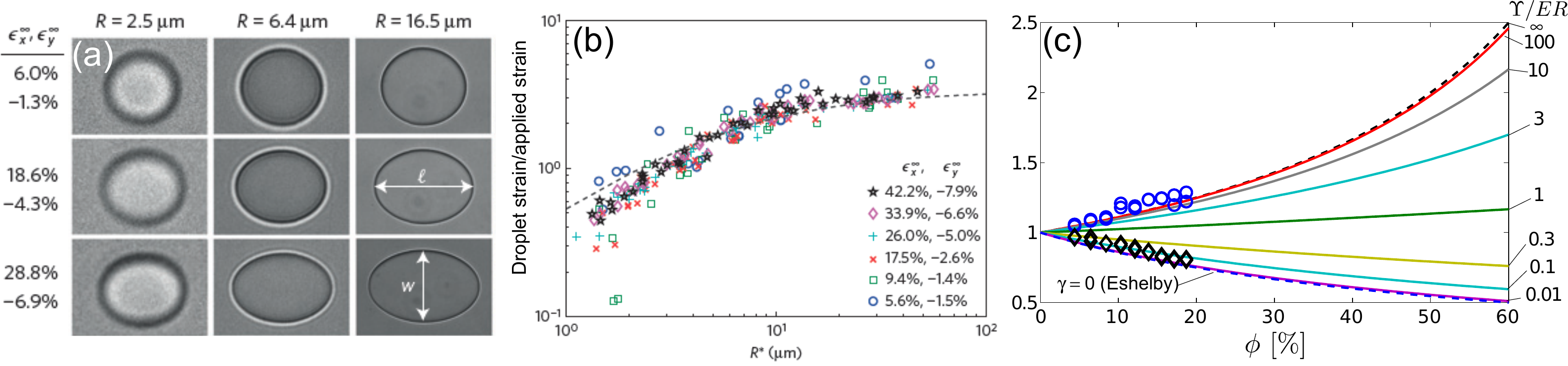} \\
    \caption{Surface stress affects the shape of small fluid inclusions in a stretched soft solid. a) Ionic-liquid droplets of different initial radii, $R$, 
    are embedded in a soft silicone sheet and stretched under plane-stress conditions. The applied $x,y$ strains are $\epsilon_x^\infty,\epsilon_y^\infty$ respectively. 
    Smaller droplets stay more spherical. Larger droplets obey classical elasticity theory. b) Droplet stretch/applied stretch as a function of droplet radius for stretched 
    ionic-liquid droplets in a soft silicone gel. Above the elastocapillary length ($\sim 10\mu$m), this is approximately constant (i.e. shape is independent of size). 
    Below the elastocapillary length, droplets stretch much less than the surrounding solid. The dashed curve is the theoretical prediction. 
    c) The stiffness of silicone/glycerol composites as a function of glycerol concentration. Blue circles: silicone with $E=3$kPa, black diamonds: 
    silicone with $E=100$kPa. In the former, increasing glycerol volume fraction, $\phi$, stiffens the composite. Curves: theoretical predictions for composites containing 
    uniformly sized droplets \cite{styl15}.} 
    \label{fig:esh}
\end{figure*}

Surface stresses stabilize spherical fluid inclusions in soft solids.
When a soft gel with a dilute concentration of embedded liquid droplets \cite{styl15} is strained macroscopically, 
the embedded droplets also deform.
However, the extent of deformation depends on the size of the droplets.
\textbf{Figure \ref{fig:esh}a} shows three inclusions of different initial sizes subjected to increasing far-field strains.
At the same imposed strains, the smaller the inclusion, the more spherical it remains.
For a given macroscopic deformation, the droplet strain varies smoothly with its radius, \textbf{Figure \ref{fig:esh}b}.
When $\Upsilon/ R E \ll 1$, the droplet shape does not depend on its size, in agreement with Eshelby's classical theory of inclusions in an elastic matrix \cite{eshe57}.
For $\Upsilon/ R E \gg 1$, the droplet shape scales with its size, as surface stresses oppose droplet deformation, contradicting Eshelby.

The stiffening of fluid inclusions by surface tension can have a dramatic impact on the macroscopic behavior of a fluid-solid composite.
Intuition and classic composite theory tell us that if you take a solid and fill a volume fraction $\phi$ of it with fluid-filled holes of radius $R$,
its effective modulus, $E_c$ will as reduce $\phi$ increases.
This works robustly when for $\Upsilon/ R E \ll 1$.
However, the opposite is observed when $\Upsilon/ R E \gg 1$.
In this case, fluid inclusions actually stiffen the composite solid (\textbf{Figure \ref{fig:esh}c}) \cite{ducl14,styl15}.

The reason that classic composite mechanics does not work is that it omits $\mathbf{\Upsilon}$, which typically acts to keep inclusions spherical.
To address this, much recent work has focussed on augmenting composite mechanics to account for surface stresses (e.g. \cite{shar04,yang04,bris10,duan05,duan07a,ducl14,styl15b}).
The augmented theory shows good agreement with experiments on emulsion \cite{ducl14} and silicone-gel \cite{styl15} composites 
when $\Upsilon$ is taken as an isotropic, strain-independent surface stress (\textbf{Figure \ref{fig:esh}b}).
The theoretical work again again highlights the importance of $\Upsilon/E$.
For example, in incompressible elastic solids containing identical incompressible fluid `holes', surface stress becomes important when $R\lesssim 100 \Upsilon/E$, 
causing stiffening relative to classical predictions \cite{styl15b}.

A practical simplification for composite calculations is approximating soft inclusions as surface-stress-free, elastic inclusions.
For example, in incompressible composites with uniform, isotropic surface stress $\Upsilon$, liquid inclusions effectively behave like elastic, 
surface-stress free inclusions with modulus $E_{\mathrm{eff}}/E=24/(9+10RE/\Upsilon)$ \cite{ducl14,styl15,styl15b}.
Thus large inclusions have vanishing effective modulus, small inclusions appear as elastic inclusions with constant stiffness 8/3 times stiffer than the surrounding matrix, 
and inclusions with $R=3 \Upsilon/2E$ are `cloaked' with $E_{\mathrm{eff}}=E$.
Making this substitution allows us to use the full power of classical composite mechanics to predict the diverse behaviour of soft composites.

\subsection{Surface Stresses Resist Fracture}

Surface stresses will also affect the opening of a crack in a soft material.
The movement of a crack tip through a solid is governed by the \emph{work of fracture} $\Gamma$, the energy required per unit area of new crack.
As this is a material parameter, we can form a second material length scale, an elasto-adhesive  length, $\Gamma/E$, which characterizes the 
radius of curvature of a crack tip at propagation (in linear-elastic fracture mechanics) \cite{cret16}.
Surface stress will then induce a closing stress on the crack tip $\sigma\sim  \Upsilon {\cal K}\sim \Upsilon E/\Gamma\sim E\epsilon$.
Thus a positive surface stress will significantly blunt a crack tip when $\Upsilon$ is comparable to $\Gamma$.
In this case crack growth will be retarded, as pointed out recently \cite{liu14,hui2016does}.

\section{ELASTOCAPILLARITY OF THREE-PHASE SYSTEMS }

So far, we have considered the impact of surface stresses on the mechanics of a single two-phase interface.
In this section, we will consider the structure of three-phase \emph{contact lines} when two or more of the phases are soft.
First, we will consider the wetting of droplets on soft solids.
Then, we will consider adhesion of rigid particles to soft solids.

\subsection{Partial Wetting on Soft Substrates}

\begin{figure*}[htbp]
    \centering
    \includegraphics[width=15cm]{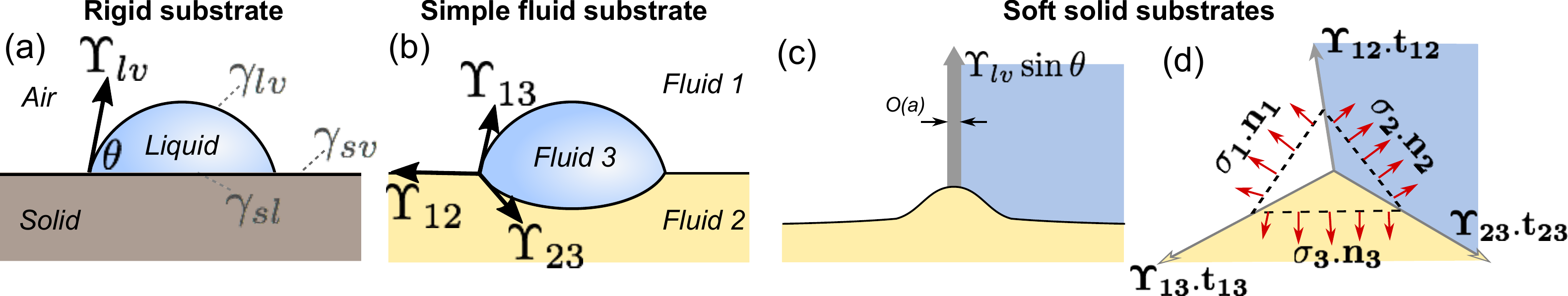}
    \caption{a) Young-Dupr\'{e}'s law: the contact angle $\theta$ is determined by the three surface energies. 
    b) Neumann's triangle: the angles at the contact line between three simple fluids are determined by a 
    force balance of the three surface stresses. 
    c) The nanoscale view of the contact line: the surface stress  is spread out over a region 
        of the order of a molecular diameter, $a$. The resulting pressure pulls the underlying surface up into a ridge. 
        d) The microscale view of a contact line on a soft solid, showing the forces acting on a small volume (with size $\gg a$) around the contact line.}
            \label{fig:wet}
\end{figure*}

There are two key results of classical wetting theory.
First, Young \cite{youn05} showed that a droplet's contact angle, $\theta$, on a rigid solid substrate was 
independent of any far-field boundary conditions (\emph{e.g.} droplet size or the thickness of a substrate), 
but varied for different combinations of materials (\textbf{Figure \ref{fig:wet}a}).
Minimising the interfacial energy of the droplet/substrate system, one finds that $\theta$  depends only on the 
surface energies of the three interfaces, $\gamma_{lv}, \gamma_{sv}, \gamma_{sl}$, 
through the Young-Dupr\'{e} relation \cite{dupr69} :
 \begin{equation}
\gamma_{lv} \cos \theta = \gamma_{sv}-\gamma_{sl}.
\label{eqn:YD}
\end{equation}

Second, Neumann \cite{neum94} considered a droplet resting on a liquid substrate with which it is immiscible, (\textbf{Figure \ref{fig:wet}b}).
At the contact line, the angles between all three interfaces are independent of the 
far-field boundary conditions and determined by a vector balance of the surface stresses:
\begin{equation}
\mathbf{\Upsilon_{12}}\cdot \mathbf{{t}}_{12} + \mathbf{\Upsilon_{13}}\cdot \mathbf{{t}}_{13} + \mathbf{\Upsilon_{23}}\cdot\mathbf{{t}}_{23} 
=0,
\label{eqn:Neumann}
\end{equation}
where the surface stress tensor and the tangent vector of the interface between phases $i$ and $j$ are denoted by $\mathbf{\Upsilon}_{ij}$ and $\mathbf{t_{ij}}$ respectively.
Over the last 100 years, equations (\ref{eqn:YD},\ref{eqn:Neumann}) have served as a basis for understanding the static structure and dynamics of fluids at small scales \cite{maxw78,dege10}.
Recently, however, it has become apparent that wetting on soft solids does not fall simply into either one of these limits.

The complexity arises because a droplet's surface tension can deform soft solid substrates.  For instance,
the out-of-plane component of a droplet's surface tension, $\Upsilon_{lv}\sin \theta$, pulls up on the surface. 
To estimate the magnitude of this deformation, let us consider a straight contact line on a semi-infinite, 
elastic substrate (\textbf{Figure \ref{fig:wet}c}).
The line-force, $\Upsilon_{lv}\sin\theta$, is spread out over an width of the order of the molecular size, 
$a$ \cite{shan86,shan87,rusa75,rusa78,whit03}.
The tensile traction applied to the substrate at the contact line can thus be approximated as 
$\Upsilon_{lv} / a$ \cite{hui14}.
This produces a strain under the contact line, $\epsilon \approx \Upsilon_{lv}/ Ea$.
When this parameter is very small,  the substrate is effectively rigid, and Young-Dupr\'{e}'s relation holds 
\cite{marc12,hui14}.

On the other hand, when $\Upsilon_{lv}/E \gtrsim a$, there are significant deformations at the contact line 
\cite{peri08,peri09,jeri11,styl13,park14}.
In this case, we can determine the contact-line geometry by considering the force balance 
on a small test volume around the contact line 
(\textbf{Figure \ref{fig:wet}d}). 
Equilibrium requires force balance between the bulk and surface stresses:
\begin{eqnarray}
\int_{W_1}\boldsymbol{\sigma_1}.\mathbf{n_1}dL+\int_{W_2}\boldsymbol{\sigma_2}.\mathbf{n_2}dL+\int_{W_3}\boldsymbol{\sigma_3}.\mathbf{n_3}dL \nonumber \\
+ \mathbf{\Upsilon_{12}.t_{12}+ \mathbf{\Upsilon_{13}.t_{13}}+ \mathbf{\Upsilon_{23}.t_{23}}=0}.
\label{eqn:fb_cl}
\end{eqnarray}
Now shrink the size of the test volume.
If the bulk stresses are bounded, or diverge more slowly than $1/r$ (which we expect), their contributions vanish while those from 
the surface stresses remain finite.
In other words, Equation (\ref{eqn:fb_cl}) reduces to Neumann's vector balance, Eq. (\ref{eqn:Neumann}), which requires the three interfaces to meet with fixed orientations  (\textbf{Figure \ref{fig:wet}b}) \cite{oliv10,styl12,styl13,hui14,lubb14,bost14}.
Note that this analysis ignores any long-range forces between the interfaces \cite{weij13,weij14}.

Thus, microscopic  behaviour near the contact line depends critically on the parameter $\Upsilon_{lv}/E a$.
Using the surface tension and molecular dimensions of liquid water, this parameter is of order one for substrates 
with Young's modulus of 100 MPa.
On much stiffer solids, such as structural materials, Young-Dupr\'{e}'s law is recovered.
On much softer solids, such as gels, Neumann's triangle describes the contact line geometry.
There is a smooth transition between the two limits \cite{leon11,marc12b,hui14,lubb14,cao15}.
This phenomenon allowed one of the first techniques for the measurement of the absolute values of solid surface stresses,
by measuring the angles between the phases at the contact line, 
and the liquid-vapour surface tension (e.g. \cite{dege10}), and using equation (\ref{eqn:Neumann}) 
to calculate the solid-vapour and solid-liquid surface stresses.

\begin{figure*}[htbp]
    \centering
	\includegraphics[width=16cm]{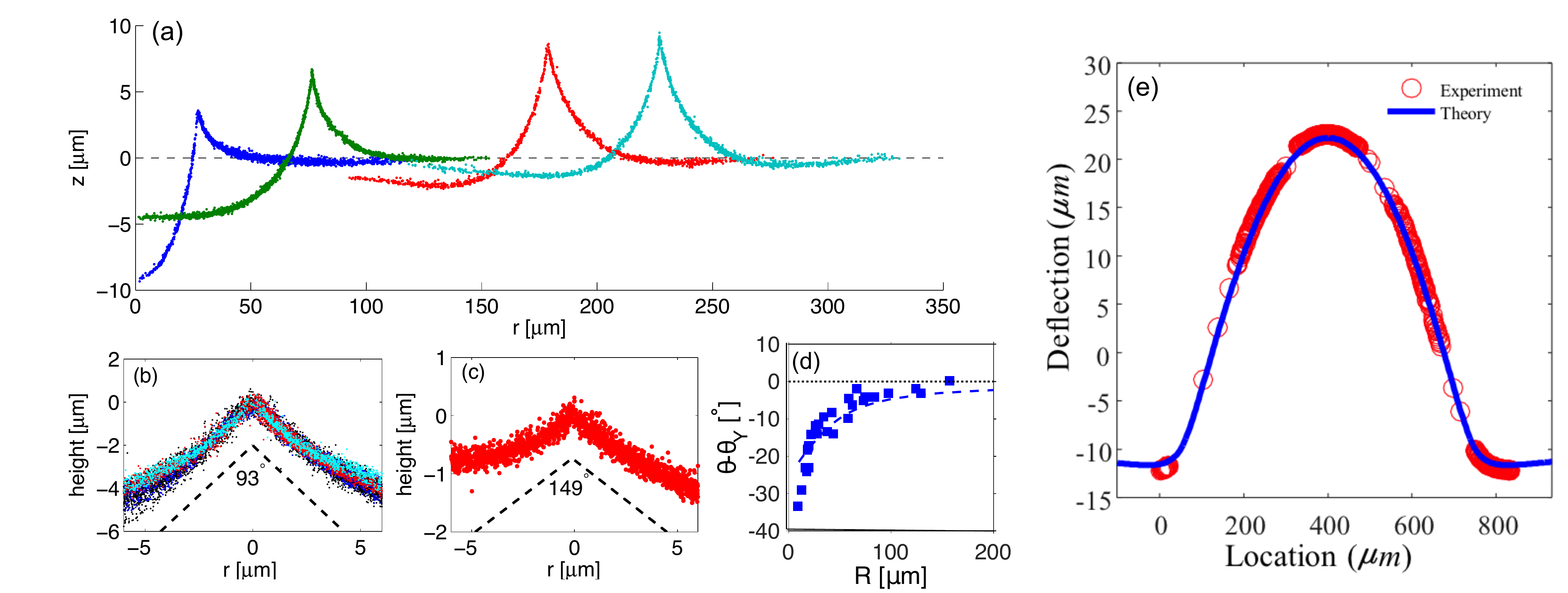}\\
    \caption{a-c) Experimental data for droplets on soft, silicone substrates ($E=3$kPa). a) Surface profiles underneath glycerol droplets of different radii. The sharp peaks correspond to the position of the contact line. b,c) when the surface profiles are aligned (by translation and rotation) at the contact line, the local ridge geometry is always found to be the same for a given liquid/substrate combination. (b: glycerol, c: fluorinated oil). d) The apparent contact angles $\theta_a$ of glycerol droplets reduce smoothly as droplet radius approaches the elastocapillary length. The dashed curve shows the linear-elastic theoretical prediction $\theta_a$ \cite{styl13b}.e) Out-of-plane deflection of a stiff silicone  film stretched across an annular disc by the Laplace pressure and surface tension of a water droplet attached to it.  The deformed shape can be represented accurately using the theory of plates with in-plane tension \cite{nade13}.}
    \label{fig:ridge}
\end{figure*}

While the shape near the contact is universal, the overall shape of the droplet depends on its size (\textbf{Figure \ref{fig:ridge}a}).
This is related to the fact that the surface stress-dominated regime near the contact line has a width of roughly $\Upsilon_s/E$, which has been confirmed experimentally 
(\textbf{Figure \ref{fig:ridge}}) \cite{styl13,park14}. 
For large droplets, $R\gg \Upsilon_{lv}/E$, the wetting ridge is small compared to the droplet size, and
the apparent contact angle is unperturbed from its value on a rigid substrate, \cite{styl12}.
Thus, large droplets behave as if they were on rigid surfaces.
On the other hand, for small droplets with $R\lesssim \Upsilon_{lv}/E$, the Laplace pressure in the 
droplet can easily depress the underlying solid surface.
In fact, for $R\ll \Upsilon_{lv}/E$, the shape of the droplet is entirely determined by the three interfacial stresses, 
as it would for a liquid substrate, (\textbf{Figure \ref{fig:wet}b}).
There is a smooth transition for the appararent contact angle from the Neumann to Young-Dupr\'{e} limits with droplet size, 
as shown in \textbf{Figure 
\ref{fig:ridge}}.

\subsection{Wetting of Slender Objects}

When a soft solid has no geometrically-imposed length scale, we have seen that  liquid surface tension can significantly deform it at scales $\lesssim\Upsilon/E$. 
However, surface tension can also deform thin rods or sheets made from stiffer materials (\emph{cf} the recent review \cite{roma10}).
In such slender geometries, elastocapillary effects can be significant at length scales much larger than $\Upsilon/E$.
A sessile droplet of characteristic size $R$ can signficantly deform  a free plate or rod when $R\gtrsim \sqrt{K_b/\gamma_{lv}}$, where
 $K_b$ is the bending stiffness of the plate/rod (in Joules).
 The length scale, $\sqrt{K_b/\gamma_{lv}}$, is also commonly referred to as an elastocapillary length.
To avoid confusion with $\Upsilon/E$, we refer to it as a \emph{bendocapillary} length, since it describes bending due to capillary forces.
Note that $\sqrt{K_b/\gamma_{lv}}$ is not a material parameter, like $\Upsilon/E$, since it depends on the 
cross-sectional dimensions of the rod or sheet.
Furthermore, bendocapillary phenomena become more pronounced as the system gets larger, in stark contrast to the 
elastocapillary deformations described in the previous section, where capillary phenomena are more pronounced for smaller droplets.
Bendocapillarity phenomena are a subject of intense research.
Examples include capillary origami, the bundling of wet hair and fibres, and clumping of micro-beams in 
micro-electromechanical devices \cite{mastrangelo1993,bico2004,kim2006capillary,py07,dupr12,wei2015elastocapillary}.
An interesting new limit has also recently been identified for the interactions of droplets with extremely bendable sheets 
\cite{huan07,davi11}.

What happens when a sheet carries tension $T$ as well as bending rigidity?  
The tension in the sheet has contributions from the surface stress on the two sides, 
$\sim 2\Upsilon$, and the bulk elastic stress $\sim \epsilon E h$, where $\epsilon$ is the tensile strain.
For very thin sheets, $h \ll 2 \Upsilon/\epsilon E$, the surface stresses dominate, and $T \approx 2\Upsilon$. 
For elastic sheets, the F\"{o}ppl-von-Karman equations predict the response of the sheet to an applied load. 
For instance, a small
drop placed on a sheet applies a line force due to liquid-vapor surface tension and distributed Laplace pressure.
By examining the magnitude of the forces due to bending and tension of the sheet, we find a new length scale $\sqrt{K_b/\Upsilon}$.
Like $\Upsilon/E$ for a thick elastic solid, this length scale captures the intrinsic deformability of a slender elastic object.
Because $K_b\sim E h^3$, it does not scale with overall system size, but only with the thickness of the sheet. 
For lateral dimensions much smaller than $\sqrt{K_b/\Upsilon}$, bending rigidity resists applied forces.
On length scales much larger than $\sqrt{K_b/\Upsilon}$, the sheet's surface stresses resists applied forces.

Thus, at length scales intermediate between $\sqrt{K_b/\Upsilon}$ and the droplet radius, $R$, 
the equilibrium configuration of the sheet near the contact line should be given by balance of tensions: a sort of Neumann's triangle, 
\textbf{Figure \ref{fig:wet}b}, with the liquid-vapour surface tension balancing the tensions on the wet and dry segments of the sheet.
Neumann's construction implies that  that significant deformations of the sheet occur when $\gamma_{lv} \gtrsim \Upsilon$.  
Since the liquid surface tension is comparable in magnitude to the surface stress, 
large deformations are generally expected for sheets in the thin limit.
As an example, Figure \textbf{\ref{fig:ridge}e} shows out-of-plane deflection of a silicone film stretched across an annular disc due to the Laplace pressure and surface tension of a water droplet placed under it.  The deformed shape can be represented accurately using plate theory \cite{nade13}.
At the macroscale, the overall shape at larger length scales resembles a droplet at a fluid-fluid interface: it obeys Neumann's triangle.
Bending is influential only in a small region near the contact edge.

These simple scaling arguments results are supported by recent theory and experiments with droplets placed on thin silicone sheets 
\cite{nade13,hui2015a,schu15}.
Thick thick/stiff sheets with $h\gg \Upsilon/E$ are undeformed by droplets and thus follow classical wetting behaviour, 
agreeing with Young-Dupr\'{e}'s law.
Thin sheets with $h\ll \Upsilon/E$ deform significantly as the Laplace pressure in the droplet causes the film to bulge out, 
taking a shape identical to fluid-on-fluid wetting.
Note that this transition does not depend on the droplet size, so the experiments can be performed with macroscopic droplets.
The contact angles of the bulging droplets can then be readily measured and used to extract the surface stresses of the sheet, 
both inside and outside the droplet (e.g. \textbf{Figure \ref{fig:ridge}e} \cite{nade13}.

\subsection{Adhesion}\label{s:Adhesion}

In adhesion theory, as with wetting, there are two key fundamental results.
First, Hertz calculated the force-indentation relationship between two adhesion-less, elastic spheres \cite{hert82}.
This forms the basis of contact mechanics, and works very well for hard materials.
Second, Johnson, Kendall \& Roberts (JKR, \cite{john71}) noticed that Hertz's theory breaks down on soft materials -- 
in particular it cannot explain why there is a significant pull-off force.
This is due to adhesion between the surfaces, typically represented by a work of adhesion,
$W$, the energy reduction per unit area of adhered surface (caused by attractive intermolecular forces).
For the case of a rigid sphere adhering  to a soft substrate, JKR showed that adhesion is important 
whenever the elasto-adhesion length $W/E$ approaches or exceeds the particle radius, $R$. 
\emph{i.e.} Hertz $\rightarrow$ JKR when $W/E \gtrsim R$.
JKR has been verified
experimentally \cite{chau91} and is widely acknowledged as the 
standard model for adhesive contact \cite{shull2002contact}.
Notably, although it assumes small contact and linear elasticity, it
is surprisingly accurate for contact radii, $a$, up to $\approx R/2$ \cite{lin2006effect}.

\begin{figure*}[htbp]
    \centering
	\includegraphics[width=10cm]{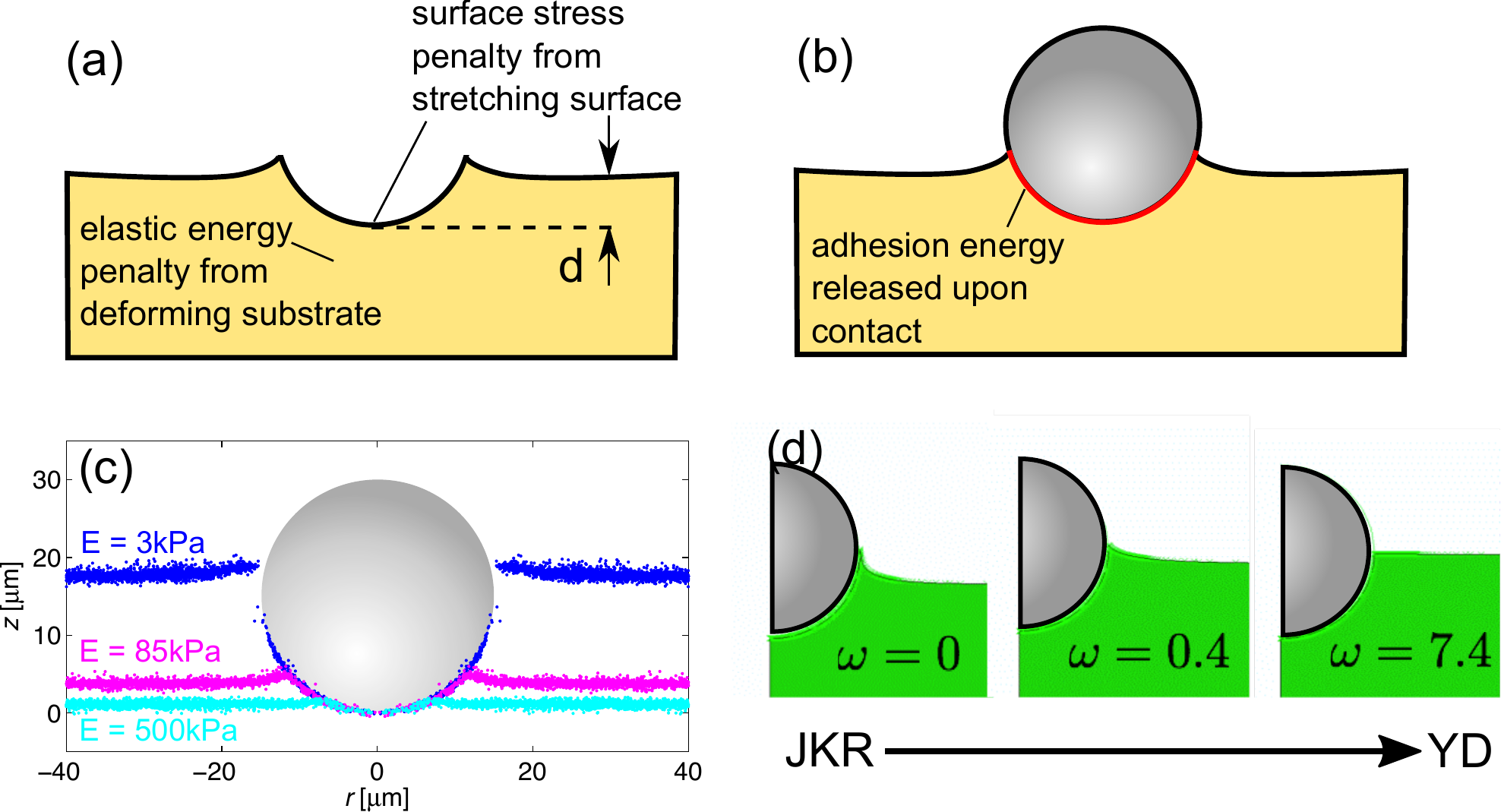}\\
    \caption{Adhesion of load-free spheres on soft substrates. a,b) Schematic of the adhesion process of a sphere to a soft substrate. c) Experimental data for glass microspheres on silicones with three different stiffnesses, measured with confocal microscopy \cite{styl13c}. d) Computed deformed shapes of a substrate in adhesive contact with a rigid sphere under zero external load \cite{xu14}.  When the adhesion parameter $\omega=0$, the deformed shape obeys JKR theory \cite{john71}.  When $\omega \gg 1$ the particle behaves as if at a fluid-fluid interface with a contact angle given by the Young-Dupr\'{e} Equation (\ref{eqn:YD}).}
    \label{fig:adhesion}
\end{figure*}

Recent experiments have shown that JKR theory breaks down on very soft solids \cite{styl13c,chak13}.
For example, load-free glass microspheres were found to indent significantly less into soft silicone gels 
than would be expected \cite{styl13c}).
This breakdown is because JKR theory neglects the role of surface stress.
To see this, consider the process of placing a load-free sphere on a soft substrate.
We break the adhesion process into two steps.
First, we deform the substrate to its final geometry (\textbf{Figure \ref{fig:adhesion}a}).
Second, we adhere the sphere (\textbf{Figure \ref{fig:adhesion}b}).
JKR theory assumes that in step 1, we only expend energy in elastically deforming the substrate, 
and this energy comes from the adhesion energy released in step 2.
However, as \textbf{Figure \ref{fig:adhesion}a} illustrates, we have significantly stretched the surface in step 1 -- 
meaning that we have also expended energy on surface work.

When will surface stress become important?
This can be determined for load-free adhesion by evaluating the work done in stretching the surface, 
$E_\Upsilon$ and the energy released upon adhesion $E_W$.
\textbf{Figure \ref{fig:adhesion}c} shows data for load-free, glass microspheres placed on soft silicone substrates.
To good approximation, the indentation takes the form of a spherical cap in a flat surface.
Thus the extra surface area created (by stretching) in the first step $\sim d^2$ ($d$ is the indentation depth) so $E_\Upsilon\sim \Upsilon d^2$.
Similarly, the adhesion occurs over the area of the spherical cap, and we find that $E_W\sim W R d$ ($R$ is the sphere radius).
Surface stress effects are negligible when $E_\Upsilon\ll E_W$, or equivalently when $d\ll WR/\Upsilon$.
From classical JKR theory, $d\sim (W^2R/E^2)^{1/3}$, so the condition for JKR theory to hold becomes $\omega\ll 1$, 
where $\omega=(W/ER)^{2/3}\Upsilon/W$ \cite{styl13c,carrillo2012contact,xu14,cao2015contact,hui2015indentation}.
This is illustrated in \textbf{Figure \ref{fig:adhesion}d} which shows finite-element computations of zero-load, adhesive contact with a rigid spherical indenter \cite{xu14}.
When $\omega=0$, the deformed shape displays the characteristic cusp of JKR theory \cite{john71}.  In the limit of large elastocapillary number the surface stress dominates and so the free surface is horizontal and meets the particle surface at an angle given by the Young-Dupr\'{e} Equation (\ref{eqn:YD}).

In fact, the dimensionless parameter $\omega$ completely determines the adhesion behaviour, even when a force  is applied to the indenter \cite{hui2015indentation}.
When $\omega \gg 1$ surface stress dominates elasticity (just as in the small droplet on a soft substrate limit).
In the limit $\omega\rightarrow 0$, JKR predicts that the pull-off force of a rigid spherical indenter from a 
soft, flat, adhesive substrate is $F_{po}=-3\pi WR/2$.
As $\omega\rightarrow\infty$, this pull-off force can be reduced by more than 30\% \cite{hui2015indentation}.
Useful, semi-analytical expressions \cite{hui2015indentation} and scaling laws \cite{carr10,styl13c,cao2015contact} 
relating force to indentation can be found in recent work and some results are provided in the Supplementary Information.

Surface stress also provides an alternative way to regularize a conceptual difficulty with JKR theory, which predicts an infinite tensile stress at the contact line.
This singularity effectively reduces JKR contact mechanics to an interface fracture problem \cite{hui2016does}, which can be regularized by the use of a cohesive zone model \cite{maugis1992adhesion}.
Alternatively, recent theoretical work has predicted that for rigid indenters on soft substrates with 
isotropic surface stresses with $\Upsilon=\gamma$, 
the soft substrate meets the indenter with a contact angle given by Young-Dupr\'{e}'s law \cite{styl13c, xu14, karp15b}. 
However, some molecular dynamics simulations disagree with this -- and observe that the contact angle is 
dependent on parameters including indenter size and substrate stiffness \cite{cao14}.
Recent experimental work has demonstrated an alternate mechanism to cut off the elastic stress singularity.
Many of the soft solids which demonstrate strong elastocapillary phenomena consist of elastic networks swollen by a solvent.
When rigid particles adhere to such a solid, the elastic singularity can be avoided by phase separation 
of the solvent from the elastic network near the contact line \cite{jens15,liu16}.

\section{SURFACE STRESS AND SURFACE ENERGY IN SOFT MATERIALS}\label{s:Theory}

From a continuum perspective, the interface of two materials is a two-dimensional sheet.
While it has no thickness, it can have mechanical properites that drive diverse phenomena.  
In this section, we review the relation between  two  surface properties,
the surface energy, $\gamma$, and surface stress, $\mathbf{\Upsilon}$.
After highlighting key concepts for simple fluids, complex interfaces and simple solids,  
we discuss the expected behaviour of soft solids like gels.

\subsection{Simple liquid interfaces \label{sec:sli}}

A theoretical understanding of the surface stress of simple liquid interfaces has been complete for some time \cite{gibbs1906scientific,rowlinson2013molecular}.  
Simple liquids are isotropic and their molecules rearrange freely under thermal fluctuations.
Thus, the increase in Helmholtz Free Energy (upon changing the surface area $A$) is simply
\begin{equation}
dF = \gamma dA \bigg|_{V,n,T}
\label{eq40}
\end{equation}
with a surface energy, $\gamma$, that is independent of fluid deformation or shape.
In equilibrium, the work done by surface stress upon stretching the surfaces must equal the change of surface free energy, $dF = dW$.
If we assume that every point on the surface moves from $\mathbf{x}$ to $\mathbf{x}+\mathbf{\Delta x}$, 
then the work done by the surface stress on an area surrounded by the closed curve $C$ is
\begin{equation}
dW=\oint_C \mathbf{\Delta x}\cdot \mathbf{\Upsilon}\cdot \mathbf{b} \,dl,
\label{eq40b}
\end{equation}
where   $\mathbf{b}$ is the outward surface normal as in \textbf{Figure \ref{fig:forces}}.
Since the interface is an isotropic fluid, the surface stress must be isotropic $\mathbf{\Upsilon}= \Upsilon \mathbf{I}_2$.
Then $dW = \Upsilon \oint_c \mathbf{\Delta x}\cdot \mathbf{b} \, dl = \Upsilon dA$.
Comparing this with equation (\ref{eq40}), we conclude that the surface stress and surface energy are numerically equivalent, 
$\Upsilon = \gamma$.

\subsection{Simple liquid with insoluble surfactants}\label{ss42} 
The next level of complexity is a fluid-fluid interface decorated by molecules or particles that are insoluble in the adjoining fluids.
Examples include phospholipids, particles, and polymers at interfaces.  
Because these adsorbed species are confined to the interface, their density changes as the surface is compressed or stretched.

For vanishing concentrations of adsorbed species, the surface stress is isotropic and identical to that of the bare interface, $\Upsilon^o$.
As the density increases, the surface stress, $\Upsilon$, decreases.
Conventionally, these systems are described by the surface pressure, 
$\Pi=\Upsilon^o-\Upsilon$ \cite{fuller2012complex}, which is readily measured in a Langmuir trough.

More generally, surface stresses can have isotropic and deviatoric contributions \cite{fuller2012complex,hermans2015lung}:
\begin{equation}
{\mathbf\Upsilon}=Tr\left({\mathbf\Upsilon}\right){\bf I}_2/3+{\mathbf\Upsilon^d}.
\end{equation}
The first term is the isotropic tension that opposes increase in surface area. 
The second, deviatoric, part represents in-plane shear forces.
This has many possible origins and is quantified by the study of surface rheology \cite{scriven1960dynamics,fuller2012complex}.  
For static systems at low concentrations, the surfactant molecules form a dilute, liquid-like layer on the surface that cannot 
support shear, and so ${\mathbf\Upsilon}^d=0$.
However, at higher concentrations, surfactants can form solid-like layers on surfaces with ${\mathbf\Upsilon}^d \neq 0$.
Non-zero ${\mathbf\Upsilon}^d$ can also be found in dynamic, fluid-like interfaces at 
finite rates of shear deformation \cite{hermans2015lung}.

\subsection{Simple solid interfaces}

To uncover the connection between surface energy and surface stress for a solid \cite{cammarata2008generalized}, 
we follow Cahn \cite{cahn1980surface} and distinguish between the reference or Lagrangian configuration $R$ and the current or 
deformed configuration $C$.
The surface free energy $F = \gamma_R A_R = \gamma_C A_C$ with areas $A_R$ and $A_C$ in the reference and current
configurations, respectively, and free energy densities $\gamma_R$ and $\gamma_C$ defined per unit area of the reference and 
current configurations, respectively.

Equating work performed on a system to its change in free energy (similar to the process in equations (\ref{eq40},\ref{eq40b})), we find
\begin{equation}
\label{eqn:shut_mk1}
{\mathbf\Upsilon}=\frac{\partial \gamma_R}{\partial {\boldsymbol\epsilon}^s}.
\end{equation}
(or $\Upsilon_{ij}=\partial \gamma_R/\partial \epsilon^s_{ij}$ in suffix notation) where ${\boldsymbol\epsilon}^s$ is the surface strain, which is assumed to be small.
This is a 2D version of the familiar connection of stress and free energy used in 3D mechanics.
Note that the 2D surface energy density being used is per unit area of the reference configuration.  
In most cases, it is more convenient to write the surface energy per unit area of the current configuration, $\gamma_C$, as 
that is what is constant in the liquid-like limit.
If the strain that connects the two configurations is small, then $A_C = A_R(1+Tr({{\boldsymbol{\epsilon}^s}}))$, 
so $\gamma_C = \gamma_R/\left(1+Tr({\boldsymbol{\epsilon}^s})\right)$.  
Then, equation (\ref{eqn:shut_mk1}) in terms of surface energy density in the deformed configuration becomes
\begin{equation}
\label{eqn:shut}
{\mathbf \Upsilon}= \gamma_C {\bf I}_2+\frac{\partial \gamma_C}{\partial {\boldsymbol{\epsilon}^s}}.
\end{equation}
This relationship is known as Shuttleworth's equation \cite{shut50}.

For small, elastic surface strains, we can derive a general form of the surface stress by Taylor-expanding 
$\gamma_R$ in terms of $\epsilon^s_{ij}$ (here we use suffix notation with the summation convention).
To leading order,
\begin{equation}
\gamma_R=\gamma_0+B_{ij}\epsilon^s_{ij}+\frac{1}{2}C_{ijkl}\epsilon^s_{ij}\epsilon^s_{kl}.
\end{equation}
If we also assume that the material is isotropic then $B_{ij}=\Upsilon_0 \delta_{ij}$ and $C_{ijkl}=\lambda^s\delta_{ij} \delta_{kl} +\mu^s(\delta_{ik}\delta_{jl}+\delta_{il}\delta_{jk})$. 
With these, equation (\ref{eqn:shut}) gives that, at leading order,
\begin{equation}
\Upsilon_{ij}=\Upsilon_0\delta_{ij}+\lambda^s \epsilon^s_{kk}\delta_{ij}+2\mu^s \epsilon^s_{ij}.
\label{eqn:surf_elast}
\end{equation}
This is the general linear-elastic form of the surface stress.
$\lambda^s$ and $\mu^s$ are the surface Lam\'{e} constants.
By analogy with bulk elasticity, $\mu^s$ is the surface shear modulus, and $K^s=\lambda^s+2\mu^s/3$ is 
the surface bulk modulus, or \emph{Gibbs elasticity}.  
If the surface Lam\'e constants are vanishingly small, the magnitude of surface
stress is constant but need not equal the surface free energy.

\subsection{Soft solid interfaces}

Now we put forward some simple hypotheses for the form of the surface stress and surface energy of a soft solids.  
We restrict our attention to polymer gels and elastomers, which have been the focus of experimental studies of elastocapillary phenomena.
Gels consist of cross-linked networks of polymer swollen by a solvent, while elastomers have no solvent.

In the limiting scenario of the `ideal gel,' the surface has the same structure and composition as the bulk.
Here, as a gel is deformed, solvent can move freely between the surface and bulk.
For a dilute gel, the bulk of the surface material is liquid, and thus the surface energy is approximately that of the solvent: $\gamma_C\approx \gamma_{l}$.
In these conditions,the gel will have an isotropic surface stress with magnitude 
that is identical to the surface energy \cite{hui13,lu2001dynamics}. 
This assumption appears to be consistent with many experimental studies on soft hydrogels, as decribed in the earlier sections, 
(\emph{e.g.} \cite{mora10,mora11,mora13}).

Generally, however, the polymer network will be perturbed by the presence of the surface.
For example, polymer chains could preferentially adsorb to the surface or the cross-linking density could vary near the surface.
Let's consider a simple extension of the ideal gel where a gel with bulk Lam\'{e} moduli $\lambda_1,\mu_1$ has different 
moduli $\lambda_2,\mu_2$ within a distance, $h$, from the surface.
Here, $h$ is assumed to be much smaller than other length scales in the problem, 
so we can
subsume the effects of the surface layer into a surface stress term.
Assuming the surface zone is still fully permeable to solvent,
then there is a constant contribution from the solvent's surface tension, as in the ideal gel.
However, upon stretching the bulk solid with a strain $\epsilon_{ij}$, the surface layer provides an excess surface stress, so that
\begin{equation}
\Upsilon_{ij}=\gamma_l \delta_{ij}+(\lambda_2-\lambda_1)h\epsilon_{kk}^s \delta_{ij}+2(\mu_2-\mu_1)h\epsilon^s_{ij}.
\label{eq:gel_Y}
\end{equation}
This simple model is consistent with the general expression for solid surface stress above  (\ref{eqn:surf_elast}).  
If the difference in moduli is  on the same order of magnitude as the modulus itself, and $h$ is no larger than a few tens of $nm$, then
we expect the 
strain dependent terms to be quite small compared to $\gamma_l$ for moderate strains.  
However, if the  moduli near the surface are much larger than those in the bulk, the strain-dependent terms become significant.
This may be expected when polymer chains, or another component of the system, absorb strongly to the interface. 

This, perhaps, is the simplest example one could imagine for non-trivial surface rheology of a gel, \emph{i.e.} 
$\Upsilon \ne \gamma \mathbf{I}_2$. 
It is very likely that the full menagerie of surface
rheological behavior such as seen in complex liquid systems (\ref{ss42})
can occur on the interfaces of soft solids.  
Further progress requires systematic measurements of the surface stress in soft solids.

\section{ANALYTICAL AND NUMERICAL METHODS}

Several methods for analysis of deformation of soft solids that account for surface stress have been developed and used. These include:
\begin{enumerate}
\item{Green's functions for continuum analyses of small-strain elastic deformations that incorporate the boundary condition
(eq. \ref{eq:gyl}) described in Section \ref{ss:strbc} \cite{hajji1978indentation,dervaux2015contact,hui14,hui2015indentation}. 
These are useful for developing analytical and semi-analytical
solutions, as well as scaling relationships. 
Key results are that the singularity of the stress and deformation fields at point or line loads on linear-elastic surfaces
are reduced significantly by the ability of surface stress to resist deformation.
Green's function techniques are closely related to transform methods which have proven useful in solving elastocapillarity problems.
These include Fourier transforms (2d geometries \cite{jeri11}), Hankel transforms (axisymmetric geometries \cite{styl12,bost14}), and Legendre transforms (spherical geometries \cite{styl15c}).
}
\item{General purpose computational (e.g., finite element) methods 
\cite{hena14,saksono2006finite1,mora13,jagota1998growth,hui02,javili2014unified}. These methods typically represent the role
of surface stress in a modular way, say as a surface finite element, which allows surface stress effects to be combined with
nearly any form of bulk mechanical behavior. In the Supplementary Material we describe a 2-node surface finite element for use with
nonlinear, implicit, static, finite-element simulations for including the influence of surface stress in plane stress/strain and axisymmetric models.
We provide a user element file (\verb!usurf_2n_2d_axi.f!) for use with the commercial finite element code ABAQUS(R) as well as an input file 
each for 2D (\verb!Rippled.inp!) and axisymmetric (\verb!Hole_axi.inp!) examples.}
\item{Molecular dynamics and density functional methods.  In molecular dynamics methods continuum properties such as elastic moduli and 
surface stress emerge automatically in terms of the underlying inter-particle potentials.  They have been used successfully to study
elastocapillary phenomena including contact mechanics \cite{carrillo2012contact,cao14,cao2015contact} and 
wetting \cite{cao2014elastocapillarity,cao2015polymeric}.  They are best suited to relatively small length
scales.
An alternative approach is Density Functional Theory (DFT) \cite{snoeijer2008microscopic}.
This shares some of the features of molecular simulation in that it is a microscopic calculation based on inter-particle potentials.
However, the model is solved semi-analytically neglecting fluctuations and representing densities by smooth functions.
It is in the spirit of the classical {\em Molecular Mechanics} methods \cite{rowlinson2013molecular}.
It has been used, for example, to study the structure near a wetting contact line \cite{marc12b}.
}
\end{enumerate}
In supplementary information we provide some further details about the Green's function for 3D and 2D problems, and about a finite element
implementation of a surface stress element in the commercial finite element code {\em ABAQUS}$^\circledR$ \cite{abaqusmanual}.

\section{CONCLUSIONS}

The research reviewed in this manuscript has established that interfaces in soft solids carry sufficient surface stress to strongly influence 
and sometimes to dominate mechanical phenomena.  
Collectively, we use the term \emph{elastocapillarity} to represent these phenomena.  
In many of these phenomena, the elastocapillary length, $\Upsilon/E$, defines the characteristic length scale over which surface stress
dominates over elasticity as the agent resisting (and sometimes driving) deformation.  
This length scale can be much larger for soft solids such 
as gels and elastomers because the interactions that determine moduli (chain entropic elasticity \cite{rubi03}) are substantially weaker than and disconnected from
those that determine surface energy and stress (near-neighbor intermolecular interactions \cite{dee98,dege10}).

In the simplest cases, surface stress is isotropic, homogeneous, and independent of surface strain, and this suffices to explain quantitatively many experiments.  
In general, surface stresses need not be any of these three.  
There is also an intimate connection between surface stress $\mathbf\Upsilon$ and surface free energy $\gamma$.  
This is captured by the Shuttleworth relation, which relates surface stress to 
how the surface free energy varies with surface strain. 
It can be used to develop prototypical 2D surface-stress/surface-strain relations.  
These are analogous to constitutive relations in bulk elasticity that can be derived from a 3D energy density and how it depends on strain.  
However, the idea of surface stress and more complex surface constitutive behaviour need not be limited to those described under 
thermodynamic equilibrium.  
We anticipate that soft solid interfaces will exhibit complex rheology, as observed at complex fluid interfaces.
Note also that we have assumed in this review that the interface has no bending rigidity (\emph{i.e.} it cannot support a moment).
However there are many soft interfaces, such as lipid vesicles and cell membranes, where bending rigidity is important \cite{seifert1997configurations}.
Thus a complete description of elastocapillarity may also need to incorporate this possibility \cite{helf73}.

Two-phase systems with a single interface represents the simplest class of problems where the role of surface stresses has been investigated.  
Phenomena such as the instability and bending of rods, flattening of a structured surface, and stiffening of a solid by liquid inclusions 
are some of the examples studied so far; there are certainly many others to be studied. For example, Eshelby theory (which we have seen is strongly modified by surface stress) is widely used beyond composite mechanics, for example in plasticity \cite{hutc70}, fracture mechanics \cite{rice68} and the cell mechanics \cite{zeme10,schw13}, so surface stress may play a role in these phenomena. 
In particular theory suggests that surface stress can strongly attenuate the energy release to a crack tip and thus effectively increase resistance to fracture, but this remains to be tested
experimentally.
Many soft materials exhibit plasticity, so we expect a whole range of `plastocapillary' effects, but this area is in its infancy \cite{styl15c}. 
Similarly, much biological material is soft, so there is almost certainly a range of biophysical elastocapillary phenomena to be uncovered (e.g. \cite{gonz15}).

The influence of surface stress on three-phase systems has been studied most in the context of wetting (two fluids and one solid), and contact
(two solids and one fluid).  
The departures from classical wetting behaviour on soft surfaces have many interesting applications and pose a wide range of 
questions for future research.
One example is the ability to control droplet motion.
Droplets spontaneously slide along stiffness gradients towards softer surfaces -- a process called droplet durotaxis that has 
parallels with cellular durotaxis \cite{styl13b}.
Droplets \cite{styl13b,karp16} and even solid objects \cite{chak14,chak14b}) spontaneously slide towards each other over 
homogeneous surfaces -- driven by substrate deformations. 
Further examples include changes in evaporation, condensation and droplet-nucleation rates on soft substrates 
\cite{soku10,esla11,lope12}; increases in the effectiveness of soft colloids as emulsifiers \cite{styl15c,mehr16}; the use of soft surfaces as protection against icing \cite{peti14}; control of the coffee ring effect \cite{lope13}; likely effects on nanobubble and nanodroplet formation on soft surfaces \cite{weij13b}; wetting of biological materials; and the potential for adhesion between soft surfaces by capillary bridges \cite{wexl14,li14} -- a strategy used by many insects \cite{labo15}.
There are many outstanding questions, both theoretical and experimental, still to be tackled.
For example, it has been known for some time that contact lines typically move at slower speeds on softer substrates.
This `viscoelastic braking' is causing by dissipation in the deforming wetting ridge under a moving contact line \cite{carr96}.
It allows the material properties of the substrate to dramatically affect speed, contact angle and smoothness of 
contact-line motion (stick-slip or not) \cite{chen11,styl13c,kaji13,karp15}.
Indeed recent work has suggested that contact-line motion can potentially be used as a sensitive measure of substrate rheology \cite{karp15}.
A second key area is the measurement of surface stresses: 
detailed imaging of the wetting ridge shape has been shown to be one of the first techniques for measuring absolute values of surface stresses.
There are also fundamental questions that have arisen from current research.
For example, in the case that surface stresses differ in magnitude from their corresponding surface energies, 
there can be rather counterintuitive effects on substrate strains near the contact line (\emph{e.g.} \cite{marc12,neuk14,hui2015a}).

Futher discussion of the outstanding questions in elastocapillarity can be found in \cite{andr16}.

\section*{DISCLOSURE STATEMENT}
The authors are not aware of any affiliations, memberships, funding, or financial holdings that
might be perceived as affecting the objectivity of this review. 

\section*{ACKNOWLEDGMENTS}
We acknowledge vibrant discussions with all of the members of our emerging community, especially those who participated in the Lorentz Center workshop on \emph{Capillarity of Soft Interfaces}.
We also acknowledge essential intellectual contributions from long-term collaborators in this topic, including John Wettlaufer and Manoj Chaudhury.
Work by ED and RWS was supported by the National Science Foundation (CBET-1236086). Work by AJ and CYH was supported by the U.S. Department of Energy (DOE), Office of Science, Basic Energy Sciences (BES), Division of Material Sciences and Engineering under Award (DE-FG02-07ER46463).

%


%
%

\bibliographystyle{ar-style4}

\end{document}